\documentclass[sigplan,screen,noacm,authorversion]{acmart}
\usepackage{fancyhdr}
\AtBeginDocument{\fancyhead[RO]{\sffamily\footnotesize Author's Version}}
\AtBeginDocument{\fancyhead[LE]{\sffamily\footnotesize Author's Version}}
\makeatletter
\def\blfootnote{\xdef\@thefnmark{}\@footnotetext}
\makeatother
\setcopyright{none}\renewcommand\footnotetextcopyrightpermission[1]{}

\usepackage{microtype}

\graphicspath{{assets/}} %

\usepackage[nameinlink]{cleveref} %
\makeatletter
\AtBeginDocument
 {
   \def\ltx@label#1{\cref@label{#1}}%
   \def\label@in@display@noarg#1{\cref@old@label@in@display{#1}}%
\def\label@in@mmeasure@noarg#1{%
    \begingroup%
      \measuring@false%
      \cref@old@label@in@display{#1}%
    \endgroup}%
 } %
\makeatother

\crefname{lstlisting}{Listing}{Listings}
\Crefname{lstlisting}{Listing}{Listings}
\crefname{listing}{Listing}{Listings}
\Crefname{listing}{Listing}{Listings}
\crefname{algorithm}{Algorithm}{Algorithms}
\Crefname{algorithm}{Algorithm}{Algorithms}
\Crefname{subsection}{Subsection}{Subsections}
\crefname{subsection}{Subsection}{Subsections}
\Crefname{section}{Section}{Sections}
\crefname{section}{Section}{Sections}
\crefname{figure}{Figure}{Figures}

\usepackage[nolist,nohyperlinks]{acronym} %

\usepackage{siunitx} %

\usepackage{xcolor} %

\definecolor{sourceBackground}{HTML}{deeed9}
\definecolor{sourceLine}{HTML}{3c9819}
\definecolor{persiststepnumfill}{HTML}{e1d5e7}

\definecolor{bluestate}{HTML}{CCD8E1}
\definecolor{greenstate}{HTML}{D3E8CB}

\usepackage{listings} %

\usepackage{tikz}
\newcommand*\circled[1]{\tikz[baseline=(char.base)]{\node[shape=circle,draw,inner sep=1.5pt] (char) {#1};}}

\usepackage{algorithm}
\usepackage{algpseudocode}

\usepackage{enumitem}

\setlist[itemize,1]{label=$\bullet$}
\setlist[itemize,2]{label=$\bullet$}
\setlist[itemize,3]{label=$\bullet$}
\setlist[itemize,4]{label=$\bullet$}
\setlist[itemize,5]{label=$\bullet$}
\setlist[itemize,6]{label=$\bullet$}
\setlist[itemize,7]{label=$\bullet$}
\setlist[itemize,8]{label=$\bullet$}
\setlist[itemize,9]{label=$\bullet$}
\renewlist{itemize}{itemize}{9}

\definecolor{nBlue}{RGB}{0, 68, 136}
\definecolor{nRed}{RGB}{204, 51, 17}
\definecolor{nGreen}{RGB}{17,119,51}

\acmDOI{10.1145/3735452.3735534}
\acmYear{2025}
\copyrightyear{2025}
\acmISBN{979-8-4007-1921-9/25/06}
\acmConference[LCTES '25]{Proceedings of the 26th ACM SIGPLAN/SIGBED International Conference on Languages, Compilers, and Tools for Embedded Systems}{June 16--17, 2025}{Seoul, Republic of Korea}
\acmBooktitle{Proceedings of the 26th ACM SIGPLAN/SIGBED International Conference on Languages, Compilers, and Tools for Embedded Systems (LCTES '25), June 16--17, 2025, Seoul, Republic of Korea}

\usepackage[mode=tex]{standalone}

\usepackage{mdframed}

\providecommand{\figurespath}{tikz}
\usepackage{tikz}
\usetikzlibrary{arrows,%
	            arrows.meta,%
	            bending,%
				calc,%
				chains,%
				decorations,%
				decorations.markings,%
				decorations.pathmorphing,%
				decorations.pathreplacing,%
				decorations.text,%
				fadings,%
				fit,%
				intersections,%
				matrix,%
				patterns,%
				positioning,%
				shadows,%
				shadows.blur,%
				shapes,%
				shapes.symbols,%
				shapes.multipart,%
				shapes.arrows,%
				shapes.geometric,%
				spy,%
				tikzmark,%
				}%

\tikzset{%
	ipetpin/.style = {%
	draw, circle, fill=black, minimum size=0.2cm, inner sep=0pt,
	outer sep=3pt,
	},
	ipetedge/.style = {%
		draw, thick
	},
	ipetlabel/.style = {%
		i4blue, circle, inner sep=0pt
	},
}

\begin{document}

\title{vNV-Heap: An Ownership-Based Virtually Non-Volatile Heap for Embedded Systems}

\author{Markus Elias Gerber}
\orcid{0009-0005-9089-4902}
\affiliation{
  \institution{FAU Erlangen-Nürnberg}
  \country{Germany}
}

\author{Luis Gerhorst}
\orcid{0000-0002-3401-430X}
\affiliation{%
  \institution{FAU Erlangen-Nürnberg}
  \country{Germany}
}

\author{Ishwar Mudraje}
\orcid{0009-0003-6870-7862}
\affiliation{%
  \institution{Universität des Saarlandes (UdS)}
  \country{Germany}
}

\author{Kai Vogelgesang}
\orcid{0000-0002-7633-1880}
\affiliation{%
  \institution{Universität des Saarlandes (UdS)}
  \country{Germany}
}

\author{Thorsten Herfet}
\orcid{0000-0002-3746-7638}
\affiliation{%
  \institution{Universität des Saarlandes (UdS)}
  \country{Germany}
}

\author{Peter Wägemann}
\orcid{0000-0002-3730-533X}
\affiliation{%
  \institution{FAU Erlangen-Nürnberg}
  \country{Germany}
}

\renewcommand{\shortauthors}{Gerber et al.}

\begin{abstract}
  The Internet of Batteryless Things might revolutionize our understanding of connected devices by harvesting required operational energy from the environment.
These systems come with the system-software challenge that the intermittently powered IoT devices have to checkpoint their state in non-volatile memory to later resume with this state when sufficient energy is available.
The scarce energy resources demand that only modified data is persisted before a power failure, which requires precise modification tracking.

We present \emph{vNV-Heap}, the first ownership-based \underline{\smash{v}}irtually \underline{\smash{N}}on-\underline{\smash{V}}olatile Heap for in\-ter\-mit\-tent\-ly powered systems with guaranteed power-failure resilience.
The heap exploits ownership systems, a zero-cost~(i.e., compile-time) abstraction for example implemented by Rust, to track modifications and virtualize object persistence.
To achieve power-failure resilience, our heap is designed and implemented to guarantee bounded operations by static program code analysis:
For example, the heap allows for determining a worst-case energy consumption for the operation of persisting modified and currently volatile objects.
The evaluation of our open-source implementation on an embedded hardware platform~(i.e.,~ESP32-C3) shows that using our heap abstraction is more energy efficient than existing approaches while also providing runtime guarantees by static worst-case bounds.

\end{abstract}



\maketitle

\blfootnote{© 2025 Copyright held by the owner/author(s). This is the author's version of the work. It is posted here for your personal use. Not for redistribution. The definitive version was published in LCTES '25.}

\section{Introduction}
\label{sec:introduction}

Intermittent computing has the potential to enable sustainable
\ac{IoT} devices that are privacy-friendly and low-maintenance~\cite{ahmed:2024:cacm}. To date,
most \ac{IoT} systems rely on batteries to power themselves.
However, this is not sustainable as batteries have a high
carbon footprint~\cite{DehghaniSanij2019,Peters2017} and must be
frequently replaced because their capacity wears out over
time~\cite{Han2019,Yang2018}. Having to replace batteries not only
becomes an infeasible maintenance effort in the presence of many small
devices but is also simply not possible for certain remote
applications~\cite{attarha:2024:assuresense}. Intermittent computing
solves these issues by harvesting energy from the environment~(e.g., from solar cells) and buffering it in an energy storage device~(e.g., capacitors). However,
this approach introduces challenges for application and
system developers, as the power may fail unexpectedly at any time.
Still, the device has to recover from a power failure quickly to be able
to make reasonable forward progress.

While early approaches to intermittent computing periodically create
checkpoints, other approaches use \emph{reactive} or \emph{just-in-time}
checkpointing~\cite{maeng:2017:alpaca,maeng:2018:osdi,jayakumar:2014:quickrecall,balsamo:2015:hibernus,balsamo:2016:hibernuspp}.
In these systems, the hardware notifies the \ac{OS} when energy is about to run out.
Asynchronous interrupts, which occur prior to an imminent power failure, allow the \ac{OS} to only create checkpoints when actually required.
This reactive checkpointing speeds up execution when energy is available but, in turn, requires the \ac{OS} to limit the amount of volatile state (i.e.,~\ac{RAM}, processor registers, and device state) that has to be persisted during a checkpoint.

Furthermore, embedded applications (including
on intermittent systems) have to date been severely constrained in their
memory usage~\cite{alonso2015dynamic,misra2004guidelines}, complicating
development and limiting their features. This not only affects novel
applications such as confidential \ac{AI}~\cite{mueller:2024:tinyep,warden2019tinyml,brown_d3zd3zzephyr_secure_inference_2022,terven_comprehensive_2023,dattu_espressifesp-tflite-micro_2025},
but also more traditional tasks like data compression and queues/buffers
in the networking
stack~\cite{vogelgesang_pfip_2025,schoeberl_tpip_2018}.

To solve this problem of constrained memory, \ac{NVM} has the potential to increase the amount of
available memory, therefore enabling more feature-rich embedded
applications. This is because it does not require refresh cycles and is
cheaper~\cite{koehler:2024:cmemento}. However, to date, it cannot be
efficiently used on embedded systems as simply replacing volatile
\ac{RAM} with \ac{NVM} comes with severe performance
overheads~\cite{jayakumar:2014:quickrecall} and with increased energy
demands to access data~\cite{maioli:2021:sensys}. This is the case even
with the most high-end \ac{NVM}
technologies~\cite{rabenstein:2023:nvram-only} that have yet to
trickle down from servers to embedded systems. To make efficient use of
\ac{NVM}, the \ac{OS} has to transparently use fast volatile \ac{RAM}
as a buffer to cache frequently accessed data from the larger \ac{NVM}
(referred to as \emph{swapping} in Linux). Unfortunately, the
traditional approach to implementing this with virtual memory requires
a \ac{MMU}, which is not available on the low-power processors required
for intermittent computing.

To summarize, embedded and intermittent systems require
appropriately-grained modification tracking and swapping
(i.e.,~\emph{virtual memory}) implemented without relying on \acp{MMU}.
The former is required to limit the amount of volatile data that has to
be persisted when a power failure is imminent. The latter is required to
increase the amount of (virtual)~memory available in order to allow for
more feature-rich applications. Existing hardware-agnostic solutions to
this problem either track modified state with insufficient
granularity~\cite{balsamo:2015:hibernus,balsamo:2016:hibernuspp,bhatti:2016:efficient}
or are too fine-grained, therefore inducing unacceptable memory- and
execution-time overheads~\cite{ahmed:2019:dice}. More flexible solutions
require the programmer to manually insert \emph{unchecked}
acquire/release calls for memory regions into their programs
to enable modification tracking~\cite{Sliper2019}.

To allow for predictable, safe, and adaptable modification tracking and
swapping for intermittent systems, we propose to leverage modern
programming-language features to allow for predictable resource behavior
through a safe interface that is checked by the compiler. For this, we
design an ownership-based \emph{\acf{vNV-Heap}} to replace traditional
virtual-memory systems with an alternative that addresses the
domain-specific requirements of embedded systems.

Using the programming-language concepts of borrowing and ownership, our \ac{vNV-Heap} can
virtualize resources, guarantee safe non-volatile memory usage
(thereby extending memory safety from volatile \ac{RAM} to \ac{NVM}),
precisely track modifications, offload unused objects, and limit
modified state. Our contributions are five-fold:
\begin{enumerate}[leftmargin=0.5cm,topsep=0.5em]
  \item \textbf{Ownership-based \ac{vNV-Heap}:} We propose the concept of
    \emph{ownership-based virtual memory} to bring features traditionally only
    available with \acp{MMU} to embedded systems. To the best of our
    knowledge, we are the first to propose the use of ownership-based
    resource \emph{safety} in virtually non-volatile heaps.
  \item \textbf{Adaptability:}
    We design the \ac{vNV-Heap} to adapt to the
    application requirements. Programmers can usually opt for convenience
    but still optimize and \emph{trade-off} memory- and execution-time overheads in critical sections
    whenever required.
  \item \textbf{Worst-Case Guarantees:} \ac{vNV-Heap} supports reliable
    intermittent systems with reactive checkpointing. For example, the amount of
    unsynchronized modified (dirty) state can be limited transparently to allow
    for in-time checkpointing when a power failure is pending.
  \item \textbf{Predictability-Focused Evaluation:} Our evaluation
    demonstrates that the \ac{vNV-Heap} outperforms \ac{RAM}- and \ac{NVM}-only
    solutions and effectively limits modified state to support reactive
    intermittent computing. %
  \item \textbf{Open-Source Implementation:} We publish our Rust-based implementation as
    open-source software and thoroughly document its structure to support future research.
    \ac{vNV-Heap}'s source code and artifact evaluation~\cite{gerber:2025:lctes:code}:\\
    \url{https://gitos.rrze.fau.de/i4/openaccess/vnv-heap}
\end{enumerate}

\section{Background}
\label{sec:background}

This section presents the intermittent systems we focus on, as well as the
relevant aspects of ownership-based memory management we rely on.

\subsection{Intermittent Computing}
\label{sec:intermittent-computing}

Traditional embedded systems assume a continuous power supply and,
therefore, require batteries to buffer large amounts of energy. However,
using batteries has significant drawbacks: They have a high carbon
footprint~\cite{DehghaniSanij2019},
their capacity wears out
over time~\cite{Han2019,Yang2018} inducing their frequent replacement,
which is labor-intensive and sometimes not feasible at
all~\cite{Lucia2017,attarha:2024:assuresense}. To solve this,
intermittently powered systems instead harvest energy from their
environment and use more durable energy buffers such as capacitors.

However, this means that intermittent systems cannot rely on a
continuous power supply. To still achieve forward progress, they must
efficiently persist and restore volatile state to retain it over power
failures. To persist their state, these systems create checkpoints at
statically determined points, dynamically at a fixed frequency, or
reactively when the charge of capacitors reaches a critical
point~\cite{Lucia2017}. Because power is an extremely critical resource
for these systems, they usually have limited memory capacity and
low-power processors. The former is the case because \ac{RAM} has a
constant power draw proportional to its capacity and is high-cost
compared to \ac{NVM}~\cite{Meena2014}. However, low-powered embedded
processors usually do not provide advanced hardware features such as an
\ac{MMU}~\cite{cortex-m-for-beginners}. These can, therefore, not be used
for augmenting the virtual memory capacity using swapping.
Any system that attempts to enable virtual-memory features, such as
swapping and modification tracking on intermittent systems, must
therefore be hardware-agnostic and efficiently exploit \ac{NVM} to
augment the available memory with minimal performance impacts.
Enabling these features allows for more feature-rich and reliable intermittent systems.

\subsection{Ownership-Based Memory Management}
\label{sec:ownership-based-memory-management}

Traditional systems programming languages such as C or C++ place the
burden of safe memory management on the programmer. Although this allows
for an easy-to-implement compiler that, in turn, allows for predictable and
high-per\-for\-mance programs~\cite{Pereira2021,Caballero2012}, it is not a
good fit for safety-critical systems as developers can easily introduce
bugs~\cite{Xu2015}~(e.g.,~use-after-free). To prevent these bugs while
still allowing for high performance and predictability, modern languages
instead rely on ownership-based memory management.

Ownership types allow for predictable, safe, and zero-cost resource management~\cite{Ferdowsi2023,clarke:1998:ownership,fahndrich:2002:lineartypes,clarke_ownership_2013}.
Rust is the most well-known example of a programming language supporting
ownership types, which gains relevance and admiration~\cite{stackoverflow:2024:developer-report}.
However, other programming languages such as Cyclone,
Ada, and SPARK also support ownership
types~\cite{Maalej2018,AdaCoreTechReport}. Specifically, in Rust's
ownership model, every value has exactly one
owner~\cite{Klabnik2023,Ferdowsi2023}. When an owned value goes out of
scope, the borrowing rules guarantee that the value cannot be
referenced, and therefore accessed, anymore. Therefore, associated
resources can be safely and predictably released (e.g.,~allocations
freed, mutexes unlocked)~\cite{Klabnik2023,Ferdowsi2023}.

\begin{figure}
\vspace{2mm}

\noindent\begin{minipage}[b]{.47\linewidth}
\centering\textbf{(a) Unsafe}

\begin{lstlisting}
(*@\setnoerror@*)mutex.lock();
*ptr += 1; // valid
mutex.unlock();
*ptr += 1; (*@\seterror@*)// race condition(*@\setnoerror@*)
\end{lstlisting}
\end{minipage}\hfill%
\begin{minipage}[b]{.47\linewidth}
\centering\textbf{(b) Ownership-Checked}

\begin{lstlisting}
(*@\setnoerror@*)ref = mutex.lock();
*ref += 1; // valid
drop(ref); // unlock
*ref += 1; (*@\setok@*)// compilation error(*@\setnoerror@*)
\end{lstlisting}
\end{minipage}

\captionof{lstlisting}{Updating a concurrently accessed, lock-protected
  value. Accessing the value after unlocking results \textbf{(a)} in a
  race condition for traditional systems programming languages, while
  \textbf{(b)} ownership rejects to compile this snippet.}
\label{lstlisting:lock-protected-value}

\end{figure}

In the following, we explain ownership by example by showing how it can
statically prevent a race condition.
\Cref{lstlisting:lock-protected-value} shows a race condition in C and
in an ownership-based language (e.g., Rust). In both code examples, a
concurrently accessed value is to be safely incremented by
locking a mutex. In a traditional systems programming language such as C,
it is easily possible to access the value by accident after the mutex is
unlocked, as shown in Sublisting~(a). This results in a race condition.
Although some languages, prominently C++, attempt to detect the most
common resource-sharing and use-after-free bugs statically, their checks
are inherently flawed as they cannot control pointer lifetimes. When
using the concepts of ownership and borrowing, the smart pointer
granting access to the value is owned exactly as long as the lock is
acquired. This is shown in Sublisting~(b). When dropping the pointer,
it goes out of scope, and the lock is implicitly released. Thereafter,
the compiler guarantees that no references to the value remain. For example, trying to access the value with the previous
pointer after it has been dropped is not possible, and compilers will reject the program.

While allocation management and mutexes are the most prominent use-cases
for ownership-based resource management, it also allows for more
sophisticated statically-checked access-control mechanisms. In the Rust
standard library, there also exists one example where mutable and
immutable access to a resource is controlled separately using ownership:
that of a reader-writer lock. This lock has a separate \texttt{lock()}
operation and guard type for readers, and one for writers. The first
provides read-only access, while only the other allows for read/write
access. The guards allow the implementation to be notified when
references go out of scope. Ownership and borrowing rules make sure that
applications can only modify data by acquiring a mutable reference via
the interface. Without an ownership-based interface, the application
would have to trivially prevent resource sharing by never handing out
references to the underlying data in the first place (e.g.,~by only
allowing for copy-based getters/setters to modify the
object~\cite{Coburn2012}), which would consequently induce expensive
copy operations for large objects.

A main inspiration for \ac{vNV-Heap}'s design is that
ownership/borrowing can also be used to restrict read/write access to a
resource separately (thereby enabling resource virtualization), and that
this kind of resource control is a great fit for solving the
shortcomings of existing virtual-memory systems and non-volatile heaps
on intermittently powered systems.

\section{System Model}

While our approach is beneficial to embedded systems in general, we put
a particular focus on supporting reactively checkpointed systems with
hard intermittency. This, in contrast to soft
intermittency, means the systems shut down entirely instead of only
entering a deep sleep mode once a critical energy level is
reached~\cite{wymore:2023:sensys}. This is desirable as it makes systems entirely independent of continuous power sources.
Furthermore, our intermittent systems persist their state using reactive
checkpoints once the charge of capacitors reaches a critical point. This
implies that energy is limited for creating a
checkpoint before power-off. To still function reliably, the routine to
create a checkpoint must have a bounded
\ac{WCEC}~\cite{maroun:2024:wcet,raffeck:2024:lctes} that does not
exceed the amount of energy available from the capacitor. If this
requirement is violated, the checkpoint cannot complete in time, and
volatile state is lost.

\section{Problem Statement}
\label{sec:problem-statement}

To enable powerful applications and ubiquitous intermittent computing,
embedded systems require virtual-memory systems that track modifications
and virtualize memory. However, low-power embedded systems most
importantly usually lack an \ac{MMU} to enable these features.

\paragraph*{Unavailability of MMUs}
\acp{MMU} are often used to implement memory swapping for increasing memory capacity. However, low-power platforms, as the one used for evaluation, usually do not include an
\ac{MMU}.
One prominent example is ARM's Cortex
family. There, the real-time--capable and low-power Cortex-M and
Cortex-R series microcontrollers do not have an \ac{MMU}. The Cortex-A
series microcontrollers provide an \ac{MMU} but also consume more
power~\cite{cortex-m-for-beginners}.
To make use of swapping and automatic modification 
tracking on low-end devices, a hardware-agnostic solution is required.

\paragraph*{Unpredictability of MMUs}
\acp{MMU} are not available on most embedded systems. However, even if
future hardware generations bring \acp{MMU} to even lower-powered CPUs,
they are unlikely to support a predictable execution time and
energy consumption required for intermittent systems. For instance,
\acp{MMU} commonly do not allow the \ac{OS} to efficiently limit the
amount of pages that are modified. For intermittent
systems, this is not acceptable, as they have to reliably persist all
modified objects in case of a pending power failure. They, therefore,
require a reliably enforced limit on the amount of modified state.
Furthermore, \ac{MMU}-based virtual memory does not directly support
real-time applications because pages can be swapped out transparently at
any time.
Instead, real-time applications must be modified to lock the required
pages into memory to make \ac{WCEC} and \ac{WCET} analysis feasible.
In summary, existing \ac{MMU}-based virtual memory systems lack central
features that are required in intermittent and real-time computing.

\paragraph*{Problem of Missing Memory Safety}
Existing research has already recognized the need for virtual memory
solutions that do not rely on \acp{MMU} and support embedded
systems~\cite{Verykios2019,bhatti:2016:efficient,maioli:2021:sensys,Sliper2019}.
These works, for example, use reference-counted pages with manual
acquire/release semantics for objects to update the reference
counters~\cite{Sliper2019}. However, these solutions suffer from
an unsafe interface. For example, they cannot prevent pointers to
swapped-out memory from leaking into program sections in which they are
invalid (similarly to use-after-free vulnerabilities). This allows for
hard-to-find and dangerous bugs that can corrupt system memory and checkpoints. To make applications safe and usage
feasible, an interface that enforces correct usage is
mandatory~\cite{Rebert2024,APTSMS2024}.

\section{Design}
\label{sec:design}

In this section, we present the concept of ownership-based
\acp{vNV-Heap}. They provide safe, hardware-agnostic, and predictable
virtual memory without relying on an \ac{MMU}.

\subsection{Requirements}
\label{sec:requirements}

To solve the problem outlined in \cref{sec:problem-statement}, our
design must support non-volatile memory safety, 
modification tracking, predictable checkpointing, and be
hardware-agnostic. The following subsections present these requirements
in
detail.

\subsubsection{Non-Volatile Memory Safety}

With unchecked interfaces for software-based virtual memory,
applications can easily corrupt their state using hard-to-find bugs. An
example of one of these bugs is when the application accesses an object
at its previous memory location, even though that object has already been
released and potentially swapped out. In this case, accessing the memory
location will overwrite other user or even system data (as
embedded systems frequently do not isolate user and kernel space).

Because of this, the \ac{vNV-Heap} must not put the burden of correctly
using an unsafe interface on the programmer. Our design offers a safe
interface that is checked at compile time by leveraging borrowing and ownership.
This prevents program bugs at compile time without runtime overheads.

In particular, applications must not be able to access stale memory
regions of swapped-out objects. We achieve this by restricting
applications to access objects only using an owned guard type~(smart
pointer) provided exclusively by \ac{vNV-Heap}'s interface. Once an
application requests such a reference, the runtime ensures that the
object is resident and not swapped out as long as this reference exists.
When the reference is finally released, the compiler makes sure the
reference is destroyed and cannot be used again. This enforces that
every time an application accesses an object, the object is resident and
the application accesses the correct memory address. Even for edge
cases, such as objects that store references to other objects, safety is
guaranteed at compile time without any involvement of the programmer.

\subsubsection{Virtualization}

To virtualize the amount of non-volatile memory, our system must track
and manage memory modifications and accesses. To enable this without
\acp{MMU}, we use ownership and smart pointers to enable
access management and tracking at allocation granularity. For this, an
application can only access its objects after requesting an immutable
or mutable reference through the \ac{vNV-Heap}'s interface. Immutable
references can be used to read data but not to modify it.
Requesting an immutable reference can never result in the object being
modified, which in turn means that any copy that resides in non-volatile
storage and was previously in-sync is still in-sync.

Only a mutable reference allows the application to modify the
corresponding object. Once a reference is requested, the \ac{vNV-Heap}
marks the corresponding object as potentially modified (this state is
also referred to as \textit{dirty} in some systems). The modified flag
can be cleared once the application releases the mutable reference, which
allows the \ac{vNV-Heap} to save the new object state to \ac{NVM}.

Our approach cannot only track modifications at allocation granularity
but also in a more fine-grained manner by dividing objects into smaller
chunks and tracking modifications for each chunk separately. This can be
appropriate for collection classes like arrays.

\subsubsection{Predictability}

To enable reactive checkpoints that can be \emph{reliably} created
on-demand when a power failure is imminent, the \ac{vNV-Heap} must be
able to persist all unsynchronized modified state stored in volatile
\ac{RAM} within a fixed and bounded \ac{WCEC}.

For this, the \ac{vNV-Heap} maintains an upper bound on the amount of
unsynchronized volatile application state at any point in time. Once an
application requests a new mutable reference, the \ac{vNV-Heap} checks
that the limit on unsynchronized modified data would not be exceeded if
the requested object were written to. If this is found to be the case,
the \ac{vNV-Heap} resolves the resource shortage by persisting modified
but currently unused objects.

\subsection{Interface}
\label{sec:interface}

The \ac{vNV-Heap}'s interface provides methods for object allocation and
reference retrieval.
Our design offers two abstraction levels: One uses automatically
inserted getter/setter calls, and the other uses borrow-checked smart
pointers, thereby resembling guard-based reader-writer locks that C++/Rust
developers are familiar with. The former allows developers to
conveniently use \acp{vNV-Heap}, while the latter allows for precise
resource control and performance. We focus on the latter in the
following because the first option is relatively straightforward to use and
can be implemented based on the low-level interface.

\subsubsection{Usage Example}

We first explain the guard-based interface using a simple example.
\Cref{lst:usage-example} shows how an application can allocate, access,
and modify objects in the \ac{vNV-Heap}. First, the example instantiates
a \ac{vNV-Heap} with a non-volatile storage device, an
\qty{8}{\kibi\byte} \ac{RAM} buffer, and the limit on the amount of
modified bytes. After the \ac{vNV-Heap} is created, the program
allocates an IP packet in the heap.

To access the allocated object, an immutable reference is retrieved by
calling \texttt{get\_ref()}. When the returned read-guard is dropped
explicitly or goes out of scope, through an implicit \texttt{drop()}, the
\ac{vNV-Heap} is invoked, allowing it to move the object out of \ac{RAM}
while it is not in use. Later retrieving and accessing a mutable
reference is analogous, except that this reference allows modifying
the underlying value. The read- and write-guards allow for transparent
access to the object's fields. At the end of the object's scope,
ownership tracking ensures the allocation is released.

\begin{figure}
  \begin{lstlisting}[emph={malloc,init,alloc,get\_ref,get\_mut,drop},emphstyle={\bfseries}]
(*@\setok@*)vnv_heap = init(
  storage = /dev/nvm1,
  cache = malloc(8KiB),
  max_modified_state = 2KiB
)

obj_handle: IP_PACKET = vnv_heap.alloc(...)

read_guard = get_ref(obj_handle)
s = read_guard.src_ip // valid read
read_guard.dst_ip = 192.0.2.1 (*@\setnoerror@*)// compilation error: invalid modify(*@\setok@*)
drop(read_guard) // release ownership

rw_guard = get_mut(obj_handle)
rw_guard.dst_ip = 192.0.2.1 // valid modify
c = rw_guard.checksum // valid read
drop(rw_guard) // release ownership
s = rw_guard.src_ip (*@\setnoerror@*)// compilation error: use-after-release
  \end{lstlisting}
  
  \captionof{lstlisting}{Example demonstrating the \ac{vNV-Heap}'s interface for a Rust-like programming language with an incoming network packet that is to be redirected. This illustrates the primitives of the low-level interface. When performance is not critical, users can use a convenient alternative based on getter/setter calls that are automatically inserted using procedural macros.}
  \label{lst:usage-example}

\end{figure}

\subsubsection{Operations}

This section presents the \ac{vNV-Heap}'s interface in detail. The
\ac{vNV-Heap}'s interface extends upon the interfaces offered by regular
heaps and reader/writer locks. The main operations to be used by
application developers are as follows.

\textbf{\texttt{init(storage, cache, max\_modified\_state)}:}
Initializes a \ac{vNV-Heap} and specifies the non-volatile storage
device, the RAM area used as a read/write cache, and the maximum amount
of unsynchronized modified state in the cache (also referred to as
\emph{dirty limit}).

\textbf{\texttt{alloc<T>()}:} Allocates a virtual object of type
\texttt{T} on the \ac{vNV-Heap}. The returned object handle represents
the object (which can be swapped-out or resident) and allows for the
retrieval of smart pointers, using \texttt{get\_ref()} and
\texttt{get\_mut()}, that must be used to access the object's value.

\textbf{\texttt{get\_ref(obj\_handle)}:} Requests an immutable
reference to the object. If this object is not yet resident, the
\ac{vNV-Heap} loads it into byte-addressable memory. If under memory
pressure, it can persist and unload unused objects to make more
\ac{RAM} available. While the application owns the returned
read-guard, the heap ensures that the object is swapped-in and
accessible (pinned).

\textbf{\texttt{get\_mut(obj\_handle)}:} This operation is analogous to
\texttt{get\_ref()}, except that the corresponding object is marked as
potentially modified.
This is done regardless of whether that application uses that reference for any
modification. It is the application's responsibility to obtain mutable
references only when necessary.
The returned access guard consequently allows
for ownership-checked mutable access. If this violates the limit
on the amount of modified bytes, the \ac{vNV-Heap} persists previously
modified but now unused objects before returning from
\texttt{get\_mut()}.

Furthermore, the \ac{vNV-Heap} registers a \textbf{\texttt{persist()}}
callback with the \ac{OS}, which is to be called when a power
failure is imminent. We design the \ac{vNV-Heap} to work with \acp{OS} or runtime systems that
support checkpointing of all (remaining) volatile application (e.g,
stack memory), system, and device state (e.g.,~\cite{raffeck:2024:lctes,bhatti2017harvos}).
A full power cycle consists of five steps:
\circled{1}
  Persist all modified data managed by the \acp{vNV-Heap}. The maximum
  number of bytes transferred to \ac{NVM} is limited by the
  \texttt{max\_modified\_state} parameter to \texttt{init()}.
\circled{2}
  The \ac{vNV-Heap} calls into the \ac{OS} to trigger the rest of the
  system to persist its state. After this, the reactive checkpoint is
  complete, and the \ac{OS} powers the device off.
\circled{3}
  After the power failure is resolved, the \ac{OS} restores all volatile
  data (e.g.,~stack, registers) except for the data managed by the
  \ac{vNV-Heap}.
\circled{4}
  The \ac{OS} calls into the \acp{vNV-Heap} asking them to restore all
  objects currently used by the application.
\circled{5}
  The \ac{vNV-Heap} returns from \texttt{persist()} and the application
  can be resumed.

Since \acp{vNV-Heap} provide a \ac{WCEC} for step 1, it is well-suited
for intermittent computing. Providing a \ac{WCEC} for step 2 depends on
the \ac{OS} and is beyond the scope of this work.
Existing energy-aware systems can address this topic~\cite{raffeck:2024:lctes}.

\subsection{Object States}
\label{sec:object-states}

\begin{figure}
	  \begin{standalone}%
\begin{tikzpicture}
  
  \tikzset{
    cnode/.style={black,draw=black,font=\sffamily\scriptsize,fill=gray!10,drop shadow, minimum width=0cm, minimum height=1.5em},
    non_resident/.style={cnode,fill=white!10, minimum width=1.5cm, minimum height=2.1em},
    resident/.style={cnode,fill=bluestate, minimum width=1.5cm, minimum height=2.1em},
    dirty/.style={cnode,fill=greenstate, minimum width=1.5cm, minimum height=2.1em},
    label/.style={font=\sffamily\scriptsize},
    arrow/.style={-LaTeX,draw=black,line width=0.3mm},
  }

  \def\paddingtop{3mm}
  \def\paddingbottom{3mm}

  \node[draw=white,inner sep=0mm, minimum width=3.33461in, minimum height=1.58in] (canvas) at (current page.center) {};
  \node[non_resident,fill=gray!10, anchor=north west, align=left] (non_resident) at ($(canvas.north west) + (1mm,-\paddingtop)$) {Swapped-Out};

  \node[resident, anchor=north, align=center]  (resident) at ($(canvas.north) + (0mm,-\paddingtop)$) {Resident};
  \node[resident, anchor=south, align=center]  (resident_in_use) at ($(canvas.south) + (0mm,8.5mm)$) {Resident,\\ Pinned};

  \node[dirty, anchor=south east, align=center]  (dirty_in_use) at ($(canvas.south east) + (-1mm,8.5mm)$) {Resident,\\Modified, Pinned};
  \node[dirty, anchor=north, align=center]  (dirty) at ($(dirty_in_use.north |- canvas.north east) + (0,-\paddingtop)$) {Resident,\\Modified};

  \draw[arrow] ($(resident.west) + (0mm, 1mm)$) -- ($(non_resident.east) + (0mm, 1mm)$) node [label,above,align=center,pos=0.5] {unload};
  \draw[arrow] ($(non_resident.east) + (0mm, -1mm)$) -| ($(non_resident.east)!0.5!(resident_in_use.west)$) node [label,rotate=90,above,align=center,pos=0.5,xshift=-9.5mm,yshift=-0.5mm] {\texttt{get\_ref()}} |- (resident_in_use.west) {};

  
  \draw[arrow] ($(resident_in_use.north) + (-4mm, 0mm)$) -- ($(resident.south) + (-4mm, 0mm)$) node [label,rotate=90,align=center,above,pos=0.5,xshift=-1.2mm,yshift=-0.5mm] {release};

  \draw[arrow] ($(resident.south) + (4mm, 0mm)$) -- ($(resident_in_use.north) + (4mm, 0mm)$) node [label,rotate=90,align=center,below,pos=0.5,xshift=0.0mm,yshift=0.0mm] {\texttt{get\_ref()}};

  \draw[arrow] ($(dirty_in_use.north) + (-4mm, 0mm)$) -- ($(dirty.south) + (-4mm, 0mm)$) node [label,rotate=90,align=center,above,pos=0.5,xshift=-1.2mm,yshift=-0.5mm] {release};

  \draw[arrow] ($(dirty.south) + (4mm, 0mm)$) -- ($(dirty_in_use.north) + (4mm, 0mm)$) node [label,rotate=90,align=left,pos=0.5,xshift=-0.25mm,yshift=0.0mm] {\texttt{get\_ref()},\\[2mm]\texttt{get\_mut()}};

  \draw[arrow] ($(dirty.west) + (0mm, 1mm)$) -- ($(resident.east) + (0mm, 1mm)$) node [label,above,align=center,pos=0.5] {sync};
  \draw[arrow] ($(resident.east) + (0mm, -1mm)$) -| ($(resident.east)!0.5!(dirty_in_use.west)$)  node [label,rotate=90,align=center,above,pos=0.5,xshift=-9.5mm,yshift=-0.5mm] {\texttt{get\_mut()}} |- (dirty_in_use.west) {};

  \draw[arrow] ($(non_resident.south) + (0mm, 0mm)$) |- ($(dirty_in_use.south |- canvas.south) + (0mm, 5mm)$) node [label,align=center,xshift=1mm,yshift=0mm] at ($(canvas.south) + (0mm, 3mm)$) {\texttt{get\_mut()}} -- (dirty_in_use.south);

\end{tikzpicture}
\end{standalone}%
    \caption{Possible object states in \acp{vNV-Heap}: Virtual
      objects can be swapped-out, resident, or modified. If the
      application owns any access guards, the object is pinned in
      \ac{RAM}.}
    \label{fig:object-states}
    \Description{The state diagram of objects managed by the vNV-Heap.
      Once an immutable reference is requested for a swapped out or
      resident object, the object is both resident and pinned. If the
      object is modified, it is still considered modified. Furthermore,
      after a mutable reference is requested for a swapped-out or
      resident object, the object is considered resident, pinned, and
      modified. If the reference of resident and pinned objects is
      released, they are no longer considered pinned. Furthermore,
      modified objects, which are implicitly resident, can be
      synchronized, after which they are no longer considered modified.
      Finally, resident objects that have not been modified can be
      swapped out, causing them to become non-resident.}
\end{figure}
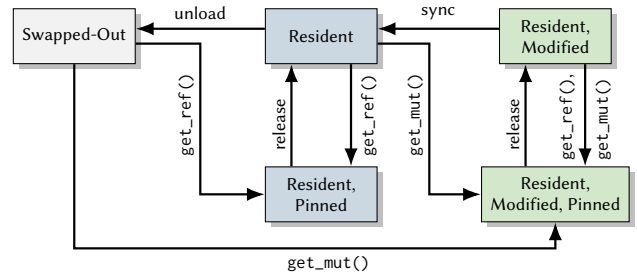

\Cref{fig:object-states} summarizes the possible states of objects
managed by the \ac{vNV-Heap}. An object can either be resident in
\ac{RAM} or swapped-out to block storage. Resident objects are either
pinned or currently inaccessible to the application. An object's
pinned marker is only cleared after the application releases the
corresponding access guard. The \ac{vNV-Heap} cannot unload pinned
objects while the application executes (unless the
application is interrupted using a signal handler as done for
\texttt{persist()}).

\subsection{Trade-Offs}

Having an ownership-checked interface enables \ac{vNV-Heap} to adapt
to the application requirements without compromising safety and
execution-time overheads. Developers can select the abstraction level
appropriate for their application and trade (unavoidable) memory- and
execution-time overheads as required. This is especially important for
real-time--capable intermittent systems, which we strive to address.

\subsubsection{Abstraction Level} Developers can either opt for a convenient
interface or optimize critical sections using our statically-checked
interface that resembles reader-writer locks. To allow for convenience
when efficiency is not required, we propose to rewrite traditional
variable uses to getter/setter calls using procedural macros at
function level. This allows programmers to write complex algorithms as
if they were using traditionally allocated objects and to automatically adapt
legacy code for common data structures. In comparison,
explicit calls to \texttt{get\_ref()} and \texttt{get\_mut()} allow the
application to speed up repeated accesses to an object by
pinning it in \ac{RAM}.

\subsubsection{Memory- and Execution-Time Overheads}
Traditional virtual memory systems usually only offer fixed
granularities at which modifications are tracked and memory-areas are
swapped in/out. Only having one page size, at which memory is managed, is
suboptimal for many applications. Hardware vendors already recognize
this problem by offering multiple page sizes
\cite{waterman_risc-v_2021,arm_translation_2025}. In comparison to
related software-based approaches~\cite{ahmed:2019:dice,Sliper2019},
\ac{vNV-Heap} intuitively allows programmers to control the granularity
by varying the size of their non-volatile allocations.

\section{Implementation}
\label{sec:implementation}

In this section, we present \ac{vNV-Heap}'s implementation. In
particular, we discuss the amount of metadata and the data structures we
use to manage allocated objects.
We implement a prototype of \ac{vNV-Heap}'s design as a generic,
modular, and platform-agnostic library. We use Rust because it fulfills
the requirements outlined in our design: First, Rust is a memory-safe
language that implements ownership~\cite{Klabnik2023}, which
\ac{vNV-Heap} uses to guarantee non-volatile memory safety. Second, Rust
is highly efficient
and provides interoperability with C and C++.

To track allocations and object states, \ac{vNV-Heap} requires 3 bytes
of metadata for each resident object (assuming a 32-bit
system)~\cite{gerber:2025:vnvheap3byte}.
This includes whether an object is modified or pinned and its (aligned)
address in \ac{NVM}.
To iterate over all resident objects, the objects and their associated
metadata is stored in a linked list. This allows the \ac{vNV-Heap} to
dynamically unload unused objects and synchronize modified objects.
Since this metadata is a part of the \ac{vNV-Heap}'s state, it also has to be persisted in case of an imminent power failure.
To provide a \ac{WCEC} for \texttt{persist()}, we count this metadata towards the application-specified limit on modified state.
To simplify our prototype, we only persist this metadata on demand during \texttt{persist()}.

\section{Evaluation}\label{sec:evaluation}

In this section, we present our predictability-focused evaluation of
\ac{vNV-Heap}.
\cref{sec:setup} shows our setup. In
\cref{sec:accessing-data}, we present the measurement-based \ac{WCET}
for \ac{vNV-Heap}'s most performance-critical primitive operation
(reference-retrieval) and compare it to a baseline implementation where
the application developer has manually split the application into
multiple parts that fit into \ac{RAM} individually.
\cref{sec:caching-effects} evaluates the speedup virtually non-volatile
objects enable by using a \ac{FIFO} queue, and
\cref{sec:persisting} demonstrates that \ac{vNV-Heap} holds its claims
on the \ac{WCEC} for the power-failure handler.
\cref{sec:dirty-tracking} compares \ac{vNV-Heap}'s performance
for a key-value store with a reimplementation of Sliper et al.'s
ManagedState~\cite{Sliper2019}. We focus on virtual persistence and
not on swapping as ManagedState only supports the former. We are not
aware of any alternative to \ac{vNV-Heap} that supports both virtual
persistence \emph{and} swapping on embedded systems.

\subsection{Setup}
\label{sec:setup}

\begin{figure}
  \vspace{2mm} %
	\centering
	\includegraphics[width=0.4\linewidth]{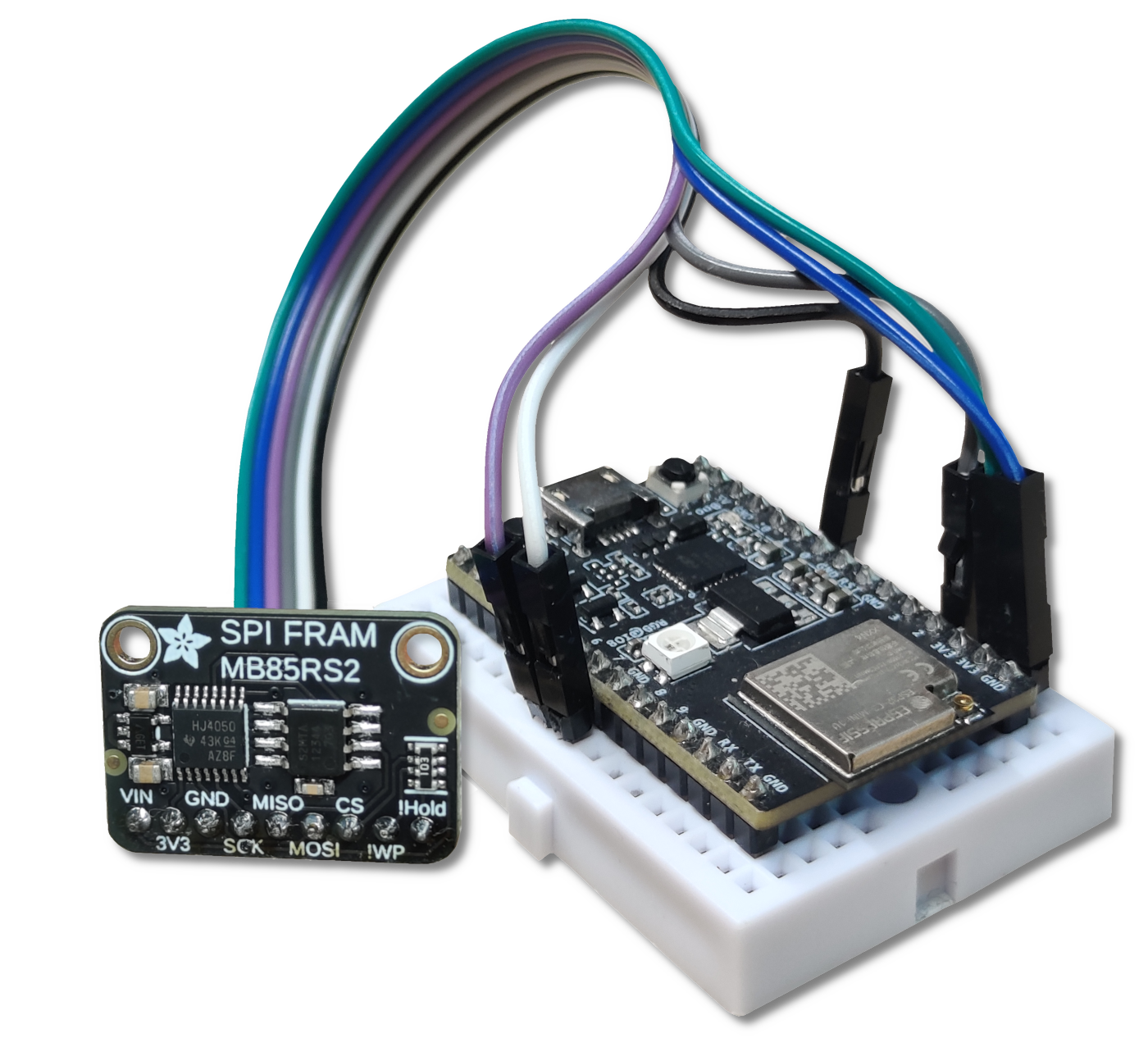}
	\caption{The evaluation system consists of an SPI-connected MB85RS4MT \acs{FRAM} module with an ESP32-C3.}
	\label{fig:esp32c3}
	\Description{The figure is fully described in its caption.}
\end{figure}

\Cref{fig:esp32c3} shows our evaluation platform. We use an ESP32-C3~\cite{ESPTechnicalReferenceManual} with a single-core
RISC-V CPU running at \qty{160}{\mega\hertz}. As \ac{NVM}, we
employ a Fujitsu \ac{FRAM} module~\cite{FRAMDatasheetNew}, which we
connect to our microcontroller via \ac{SPI} at an operating frequency of
\qty{40}{\mega\hertz}.
As \ac{RTOS},
we use Zephyr v3.7~\cite{zephyr:2024:v3.7} with
\textit{zephyr-rust}~\cite{tylerwhall:2025:zephyr-rust}.

To better emulate the performance characteristics of integrated
\ac{FRAM}, we do not exploit the fact that block writes can be sped up
significantly by issuing large bulk operations over \ac{SPI}. Instead,
we issue a separate \ac{SPI} call for every word (4 bytes) transferred.
Because our \ac{FRAM} is not efficiently byte-addressable, we also do not
exploit the fact that \ac{vNV-Heap} could allow the application to
directly modify objects stored in \ac{FRAM}.\footnote{To implement this,
\cref{fig:object-states} could be extended by allowing
\texttt{get\_ref/mut()} calls on \textit{Swapped-Out} objects.
Frequent accesses should still cause the object to be moved to
\ac{RAM}. We exclude this as we do not have the hardware needed to
evaluate that configuration available.} Both decisions are only to the
disadvantage of \ac{vNV-Heap} and do not impact the evaluation's main
results (only the \ac{NVM}'s access latency being technology-specific
anyway).
To further speed up performance, either an integrated \ac{FRAM} can be chosen, which has no \ac{SPI} overhead, or bulk access can be exploited.

We use ESP32-C3's on-board timer (running at
\qty{16}{\mega\hertz}) for execution-time measurements and disable
Zephyr's scheduler. We clear CPU caches before each measurement.

\subsection{Reference Retrieval}
\label{sec:accessing-data}
The operations used to retrieve/release mutable and immutable references are the
most performance-critical operations of the \ac{vNV-Heap}; we therefore evaluate their
performance in this section. In contrast,
allocation and deallocation usually happen much more infrequently, for example,
during program initialization.

We empirically trigger the best- and worst-case latencies of
\texttt{get\_ref()}. Furthermore, because the worst-case performance is
very unlikely to be observed in real-world applications, we also measure
a more typical worst-case latency, which we term \textbf{bad-case} latency. The
different code paths are triggered under the following conditions:

\begin{description}%
\item[Best-Case] The target object is already resident.

\item[Bad-Case] The requested object is not resident;
however, loading it into \ac{RAM} does not require unloading other
objects.

\item[Worst-Case] To trigger this, the application must
allocate objects that are as large as the entire \ac{RAM} available.
Then, this case measures the latency when a target object is not
resident and loading it into \ac{RAM} requires persisting and
unloading a maximum-sized modified object (i.e.,~one object that fills
the entire volatile cache of each system).

\end{description}

\noindent
We compare our \ac{vNV-Heap}--based application with an application that
manually switches between multiple modules that each fit into \ac{RAM}
individually. Whenever code from a module is to be executed, the
respective module is swapped in, with the currently resident module
getting swapped out to \ac{NVM}. This represents a realistic
architecture an engineer can pursue if they realize that the entire
application's code and data do not fit into the available \ac{RAM}.

In the worst-case scenario, the maximum amount of \ac{RAM} used has a
significant impact on the latency observed. We configure both
implementations to use at most \qty{1}{\kibi\byte} of \ac{RAM}, which is
then also the maximum object size and the limit for resident and
modified objects.

\begin{figure}
  \vspace{2mm} %
	\centering
	\includegraphics[trim={0 0.1in 0 0.1in},clip]{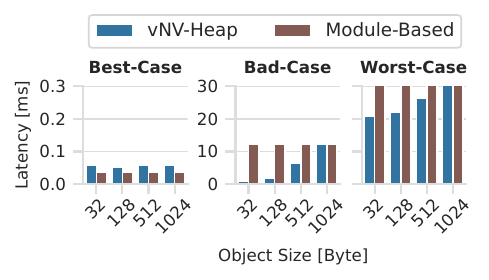} %
	\caption{Latency for accessing an object in the \ac{vNV-Heap} and with
		a module-based swapping approach (baseline). Both systems are
		configured such that the size of the largest resident object is
		\qty{1}{\kibi\byte}. Note that the scale for the best-case differs
		from the other two scales.}
	\label{fig:accessing-data}
	\Description{Three bar plots comparing the latency for accessing a single object between the vNV-Heap and a module-based swapping approach (which serves as the baseline). Each of the three plots shows this latency (on the y-axis) for four object sizes (on the x-axis), which are 32, 128, 512 and 1024 bytes. The first bar plot shows the best case and a latency of approximately 56 microseconds for the vNV-Heap, while the baseline has a latency of approximately 37 microseconds. These numbers remain relatively constant for all four object sizes. The second bar plot shows a bad case and a latency of approximately 12 milliseconds for the baseline, while the vNV-Heap improves performance by 93\%, 84\%, 48\% and 0.5\% for a 32-, 128-, 512- and 1024-byte object, respectively. The third bar plot shows the worst case and a latency of approximately 32 milliseconds for the baseline, while the vNV-Heap improves performance by 36\%, 32\%, 19\% and 0.8\% for a 32-, 128-, 512- and 1024-byte object, respectively.}
\end{figure}

\Cref{fig:accessing-data} shows the latency that the \ac{vNV-Heap} approach and the
module-based approach introduce when (virtually) non-volatile objects
are accessed. The best-case latency for the \ac{vNV-Heap} is
twice as high as the baseline. This is due to the additional management
overhead.
However, the absolute latency is still very low (below
\qty{100}{\micro\second}) and, therefore, irrelevant to the real-world
performance of most applications. In the bad-case, the \ac{vNV-Heap}, in
return, allows the application to save tens of milliseconds. It is at
least as fast as the baseline in every configuration and even reduces
the latency by 93\% for 32-byte objects and by 57\% on average.
In the worst-case, the \ac{vNV-Heap} improves performance by 22\% on average.

\subsection{Read/Write Cache}
\label{sec:caching-effects}

Applications can use \acp{vNV-Heap} to speed up reads and writes of
persistent data by using the available RAM as a cache (similarly to the
page cache in Linux). To evaluate this, we benchmark a (virtually)
non-volatile \ac{FIFO} queue, storing 256-byte objects. In
intermittent computing, such queues can be used to buffer jobs until the
energy for processing becomes available~(e.g., incoming data or network requests).

\begin{figure}
  \vspace{2mm} %
	\centering
	\includegraphics[trim={0 0.1in 0 0.1in},clip,width=\linewidth]{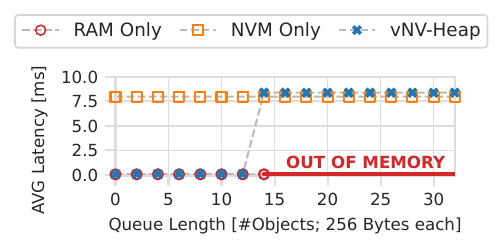} %
	\caption{The average latency for inserting one element and then removing the least recent one from a queue with a specific initial length. The volatile \ac{RAM} usable by each queue implementation is limited by \qty{4}{\kibi\byte}.}
	\label{fig:queue}
	\Description{A line plot showing the average latency for pushing and then popping one element from a queue with a given initial length. This plot compares three implementations. The first one is a RAM-only approach that stores the queue exclusively in volatile memory. Pushing and popping one element has a low and constant latency of 56 microseconds. However, this approach runs out of memory when the queue size exceeds 14 elements. The second implementation is an NVM-only approach that stores the entire queue in NVM. This has a relatively high and constant latency of 7983 microseconds for pushing and popping one element from the queue. The third implementation is powered by the vNV-Heap. For queue lengths less or equal to 12, it has a relatively low latency of 84 microseconds on average. For queues that have 14 or more elements, the latency to push and pop an element is approximately 8362 microseconds.}
\end{figure}

\Cref{fig:queue} shows the average latency for a RAM- or NVM-only queue
and for a vNV-Heap--based queue. We start with a variable number of elements
in the queue and measure the latency for inserting and removing one
final element. \ac{RAM} usage is limited to \qty{4}{\kibi\byte} in all
cases.
\begin{description}%
	\item[\ac{RAM} Only:] Fast volatile \ac{RAM} is used exclusively to store the queue.
		Because of this, the queue's length cannot exceed 15 elements.
	\item[\ac{NVM} Only:] All elements are stored in \ac{NVM}.
	\item[\ac{vNV-Heap}:] This variant combines fast volatile
    \ac{RAM} and \ac{NVM}. If possible, the \ac{vNV-Heap}
    stores the entire queue in volatile \ac{RAM}, otherwise elements are
    dynamically swapped to \ac{NVM}. Because allocation metadata
    imposes some storage overhead, the \ac{vNV-Heap} needs to unload
    elements for queues longer than 12 elements.
\end{description}
\noindent
Using the \ac{vNV-Heap}, applications can, therefore, take advantage of
\ac{RAM}'s low latency without being limited by its size.

\subsection{Predictable Checkpointing}
\label{sec:persisting}

This section shows that the \ac{vNV-Heap} can create reactive checkpoints
predictably. As the baseline, we choose a system consisting of unmanaged
memory that creates reactive checkpoints by copying all bytes in its
region to \ac{NVM}. Contrary to such a system, the \ac{vNV-Heap} reduces
the \ac{WCEC} by enforcing a limit on the amount of modified state. Both
variants are configured with the same amount of \ac{RAM}.

\begin{figure}
  \vspace{2mm} %
	\centering
	\includegraphics[trim={0 0.1in 0 0.1in},clip,width=\linewidth]{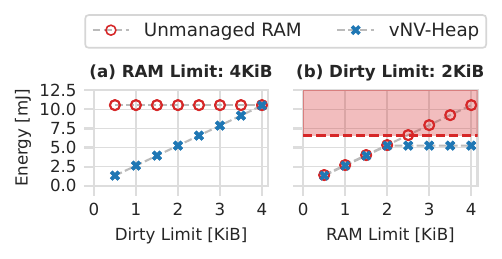} %
	\caption{The measurement-based \ac{WCEC} for persisting the \ac{vNV-Heap} compared to unmanaged \ac{RAM}.}
	\label{fig:persist}
	\Description{Figure 5 fully described in the text.}
\end{figure}

For our evaluation, we determine the checkpoint creation's \ac{WCEC}
as follows: First, we determine the measurement-based
\ac{WCET} for the \ac{vNV-Heap} and the baseline.
For this, we also measure the \ac{WCET} of all critical sections that may delay the creation of reactive checkpoints.
As examples for this delay serve active I/O operations issued just before \texttt{persist()}, since these I/O operations first have to complete.
Such I/O semantics, for example, exist with the transactional and uninterruptible behavior of \ac{SPI}.
To calculate the \ac{WCEC} from the measurement-based
\ac{WCET}, we assume that our ESP32-C3 consumes a maximum of
\qty{40}{\milli\ampere} at \qty{3.3}{\volt} (\qty{132}{\milli\watt}).
\Cref{fig:persist} displays two cases where the \ac{vNV-Heap} improves
the \ac{WCEC} relative to a traditional system.

\Cref{fig:persist} (a) assumes the system has \qty{4}{\kibi\byte} of
RAM. We show the \ac{WCEC} for different limits on the amount of
modified state (x-axis, \emph{dirty limit}). As the baseline cannot
enforce this, it must save all \qty{4}{\kibi\byte} of volatile \ac{RAM}
when creating the checkpoint. Therefore, the respective \ac{WCEC} is
fixed but suboptimal. The \ac{vNV-Heap}, in turn, already persists a
subset of its volatile data at runtime and, therefore, reduces the
\ac{WCEC} for the checkpoint.

In \Cref{fig:persist} (b), the maximum amount of modified state is fixed
to \qty{2}{\kibi\byte}, but we show the \ac{WCEC} for different
RAM sizes. The \ac{WCEC} for checkpoint creation using the
\ac{vNV-Heap} remains bounded even if the system uses more than
\qty{2}{\kibi\byte} of RAM. The baseline's \ac{WCEC}, however, increases
with the amount of volatile \ac{RAM}.

In summary, the \ac{vNV-Heap} can create reactive checkpoints
predictably by limiting the routine's \ac{WCEC}.

\subsection{Key-Value Store}
\label{sec:dirty-tracking}

In this section, we compare \ac{vNV-Heap} against ManagedState by Sliper
et al.~\cite{Sliper2019}, which is the closest alternative solution we
are aware of. Regarding performance, we focus on a typical
key-value--store workload.
ManagedState allows the system to limit and reduce the energy
consumption for creating and restoring reactive checkpoints. It manages
memory at page granularity and limits the number of pages potentially
containing modified data. Like \ac{vNV-Heap}, ManagedState is
hardware-agnostic and requires applications to explicitly request
read/write access before using a memory region (which might be a
sub-/superset of a page). Contrary to \ac{vNV-Heap},
ManagedState's interface does not enforce safety and thus allows for
dangerous use-after-release bugs. Also, data can be modified
accidentally without first requesting mutable access. Furthermore,
ManagedState does not support memory swapping and thus can only provide as much
storage as the \ac{RAM} size.

In particular, \ac{vNV-Heap} differs from ManagedState in that the
latter requires developers to select a fixed page size for each
application while \ac{vNV-Heap} manages memory at
allocation granularity. One page size might not be optimal for heterogeneous
applications. When it is chosen too large, this negatively
impacts the reactive checkpoint because whole pages have to be persisted
(even when only a portion was modified). When the page size
is chosen too small, this needlessly increases the management overhead.
Because metadata is part of the volatile system state and therefore
included in the checkpoint, small pages reduce the effective amount of
\ac{RAM} usable for the applications.
Also, applications are more likely to perform modifications at allocation granularity than at page granularity.

Unlike ManagedState, \ac{vNV-Heap} does not fix a page size globally
but intuitively adapts to the application by enforcing virtual
persistence at allocation granularity. Even though \ac{vNV-Heap}
objects may also be chosen too small or too large, developers in
comparison still benefit from \ac{vNV-Heap}'s adaptability (regarding
the metadata overhead) and enhanced feature set (i.e.,~swapping and
memory safety).

To compare both approaches, we use a custom embedded key-value store
with numbers as keys and variable-sized values. Based on Cao et al.'s
analysis of key-value stores~\cite{Cao2020}, we randomly insert
a total of 256 objects with different value sizes into the store:
64$\times$32 bytes, 128$\times$128 bytes, 32$\times$256 bytes, and
32$\times$1024 bytes (\qty{58}{\kibi\byte} in total). We also limit the
amount of modified bytes to one-fifth of the total user data. We count
both metadata and the actual user data towards this limit. We further
favor ManagedState in that all data fits into volatile \ac{RAM} (only
\ac{vNV-Heap} supports swapping). We still include all of
\ac{vNV-Heap}'s metadata in the evaluation, even if it is only required
for swapping.
For this, we assume 3 bytes per object as this would be achievable using optimizations.
To increase code readability, our currently unoptimized \ac{vNV-Heap} implementation uses 20 bytes~(including padding).

\begin{figure}
  \vspace{2mm} %
	\centering
	\includegraphics[trim={0 0.1in 0 0.1in},clip]{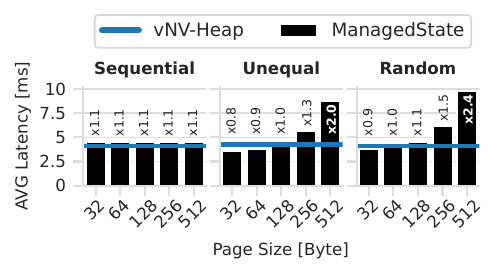} %
	\caption{The latency for updating values in a key-value store
    evaluated for three access patterns. The \textit{Unequal} access
    pattern updates a subset of values frequently, while the
    remaining values are accessed infrequently.}
	\label{fig:kvs}
	\Description{Three bar plots showing the latency for updating values in a key-value store for three access patterns. In all three subplots, the latency of ManagedState for five different page sizes (32, 64, 128, 256, and 512 bytes) is compared with vNV-Heaps. For the sequential access pattern the latency of ManagedState is relatively constant for all five page sizes. In this case, the vNV-Heap slightly improves performance by an average of 9\%. For the second access pattern, called "Unequal", the vNV-Heap adds a performance overhead of 18\%, 14\%, and 2\% for page sizes of 32, 64, and 128 bytes, respectively. However, for page sizes of 256, and 512 bytes, the vNV-Heap improves performance by 25\%, and 51\%, respectively. This is similar to the last access pattern, called "Random". Here, the vNV-Heap adds a performance overhead of 8\%, and 3\% for a page size of 32, and 64 bytes, respectively. For page sizes of 128, 256, and 512 bytes, the vNV-Heap improves performance by 8\%, 34\%, and 58\%, respectively.}
\end{figure}

\Cref{fig:kvs} compares the latency for updating whole values in both
systems. We use a Rust-based reimplementation of ManagedState for which
the source code is also available in \ac{vNV-Heap}'s git repository.
To diversify our results, we use three different access patterns to
update values. For the \textit{Unequal} access pattern, a subset of keys
is updated frequently, while the remaining keys are updated less
frequently. The non-normalized distribution to calculate access
probabilities for each key is $\sin^{4}(\frac{5}{32} \cdot \text{key}) +
0.1$.

\begin{figure}
  \vspace{2mm}
  \includegraphics[trim={0 0.1in 0 0.1in},clip]{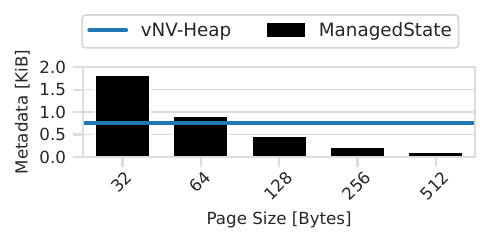} %
  \caption{The total number of bytes used for metadata in the \ac{vNV-Heap}
    and ManagedState. The state of both systems is equivalent to
    \Cref{fig:kvs}.}
  \label{fig:kvs-metdata}
  \Description{A bar plot comparing the amount of metadata required by the vNV-Heap versus ManagedState for five different page sizes (32, 64, 128, 256, and 512 bytes). For page sizes of 32, and 64 bytes, the vNV-Heap improves metadata usage by 59\%, and 17\%, respectively. For page sizes of 128, 256, and 512 bytes, the vNV-Heap introduces metadata overheads of 66\%, 231\%, and 562\%, respectively.}
\end{figure}

Even though the \ac{vNV-Heap} is more feature-rich, its performance is
comparable to ManagedState for small page sizes. For large page sizes,
\ac{vNV-Heap} clearly outperforms ManagedState. Furthermore,
\Cref{fig:kvs-metdata} shows that ManagedState's metadata overhead (1
byte per page) is significant for small page sizes. While \ac{vNV-Heap}
requires 3 bytes per object, this does include the heap-management
overhead that would have to be added for ManagedState if the application
wants to benefit from a heap abstraction. Even without that,
\ac{vNV-Heap} provides significant benefits to heterogeneous
applications for which a fixed page size is not sufficient.

\section{Related Work}
\label{sec:related-work}

We group related works into systems dependent on \acp{MMU} and
hardware-agnostic approaches. To the best of our knowledge,
\ac{vNV-Heap} is the first solution that provides a safe,
hardware-agnostic, predictable, and adaptable alternative to traditional
virtual-memory systems for intermittent systems.

\subsection{MMU-Dependent Approaches}
Most well-known approaches to virtual persistence require
\acp{MMU} and are thus not applicable to intermittent systems.

Coburn et al.'s NV-Heaps~\cite{Coburn2012} allows applications to access
persistent objects through a getter/setter-based interface. In contrast,
our \textbf{v}NV-Heaps improve upon this by being hardware-agnostic and
by augmenting the getter/setter-based interface with a more efficient,
but still safe, ownership-based interface. Only this allows for the
precise resource control and predictability that intermittent systems
require.

Both TreeSLS~\cite{Wu2023} and Aurora~\cite{Tsalapatis2021} are examples
of single-level stores. These are \acp{OS} that manage and persist the
entire system state, including the state of all applications.
Unlike \acp{vNV-Heap}, these systems take periodic checkpoints at fixed
time intervals and use the most recent checkpoint to restore the system
state after a power failure. To allow for incremental checkpoints,
TreeSLS and Aurora depend on an \ac{MMU} to track modifications at
page granularity. In contrast, \acp{vNV-Heap} work on low-power embedded
systems without an \ac{MMU} and allow the application to adapt the
granularity at which modifications are tracked based on its needs.

\subsection{Hardware-Agnostic Approaches}

The most prominent alternative to \ac{vNV-Heap} we are aware of is
ManagedState by Sliper et al.~\cite{Sliper2019}.
\cref{sec:dirty-tracking} not only compares their features but also measures how they compare for a real-world key-value store workload.

Yerkios et al.~\cite{Verykios2019} and Bhatti et
al.~\cite{bhatti:2016:efficient} propose systems to track the different
memory segments, which allows them to persist only actually used memory.
They do not track modifications and, therefore, have to assume that the
application modifies all of its accessible memory (implying an increased
checkpointing overhead).

Like \ac{vNV-Heap}, ALFRED~\cite{maioli:2021:sensys} provides virtual memory for improving energy efficiency of intermittent systems.
Thereby, ALFRED primarily targets the minimization of \ac{NVM} use.
It achieves this by deciding what data to move to volatile \ac{RAM} or to \ac{NVM} at compile time, and thereafter inserting specific instructions into the code.
An integral contrast between \ac{vNV-Heap} and ALFRED is that ALFRED has a static mapping of variables either to volatile or to non-volatile memory.
Instead of a static mapping, \ac{vNV-Heap} performs dynamic object tracking with the concept of~(statically checked) ownership and uses the fast volatile memory for reading/writing objects.
In case of imminent power failures, \ac{vNV-Heap} stores the dirty memory to persistent memory predictably, which is relevant for our scenarios of reactive intermittent computing.

While MPI~\cite{grisafi:2022:mpi} also relies on compiler-based
isolation to prevent checkpoint corruption, they do not employ ownership to elide runtime checks.

Surbatovich et al. introduced Curricle, a type system for intermittent computing~\cite{surbatovich:2023:pldi}.
Comparable to \acp{vNV-Heap}, Curricle also exploits the Rust programming language.
In contrast to Curricle, \acp{vNV-Heap} rely on WCEC analysis to ensure forward progress, which is comparable to WoCA~\cite{raffeck:2024:lctes}.
With the WCEC-based approach, \acp{vNV-Heap} does not demand users to annotate (non-)idempotence requirements of variables.

\section{Conclusion}

We have demonstrated that ownership-based \acfp{vNV-Heap}
enable more powerful yet sustainable embedded applications through
intermittent computing. These systems do not require batteries that
degrade over time but can still avoid stringent restrictions on
memory usage by extending volatile \ac{RAM} with cheaper \acf{NVM}. They
do this with minimal execution-time overheads as frequently accessed
data is cached in fast volatile \ac{RAM}. The \ac{vNV-Heap} supports
worst-case guarantees and has a safe yet convenient interface that still
allows developers to precisely control resource usage whenever required.
In summary, our proposal provides a \emph{safe}, \emph{predictable}, and
\emph{adaptable} replacement for traditional virtual-memory systems that
is tailored to the needs of modern \ac{IoT} devices.

\medskip

\begin{mdframed}[
  backgroundcolor=greenstate!15,
  linecolor=greenstate!90,linewidth=2pt,topline=false,bottomline=false,
  innertopmargin=0pt, innerbottommargin=5pt, innerleftmargin=6pt, innerrightmargin=6pt,
  frametitleaboveskip=5pt,
  frametitlefont=\itshape, frametitlealignment=\centering,
  frametitle={\ac{vNV-Heap}'s source code \& artifact evaluation~\cite{gerber:2025:lctes:code}:}]
\centering
\url{https://gitos.rrze.fau.de/i4/openaccess/vnv-heap}
\end{mdframed}

\begin{acks}
  We thank Wolfgang Schröder-Preikschat for his initial inspiration for
  this work and Dustin Nguyen for the feedback.
  This work is supported by the German Research Foundation~(DFG) --
  project number 502615015~(ResPECT) and
  project number 502947440~(Watwa).

\end{acks}

\bibliographystyle{ACM-Reference-Format}
\bibliography{references-final}

\begin{acronym}[ABCDEFGHIJK]
    \acro{FIFO}{First In, First Out}
    \acro{FRAM}{Ferroelectric Random Access Memory}
    \acro{AI}{Artificial Intelligence}

    \acro{IoT}{Internet of Things}
    \acro{NVRAM}{Non-Volatile Random-Access Memory}
    \acro{NVM}{Non-Volatile Memory}
    \acro{MMU}{Memory Management Unit}
    \acro{OS}{operating system}
    \acro{PFI}{Power-Failure Interrupt}
    \acro{RAM}{Random-Access Memory}
    \acro{RTOS}{Real Time Operating System}
    \acro{SPI}{Serial Peripheral Interface}
    \acro{SoC}{System on a Chip}
    \acro{vNV-Heap}{virtually Non-Volatile Heap}
    \acro{WCEC}{Worst-Case Energy Consumption}
    \acro{WCET}{Worst-Case Execution Time}
\end{acronym}

\end{document}